\input harvmac
\input epsf.tex
\def\caption#1{{\it
	\centerline{\vbox{\baselineskip=12pt
	\vskip.1in\hsize=5.0in\noindent{#1}\vskip.1in }}}}
\def\pyidk{PHY-9057135}
\def\dint{\int \kern-.6em \int\kern-.2em}

\def\({\left(}
\def\){\right)}
\def\[{\left[}
\def\]{\right]}
\def\CJ{{\cal J}}
\def\CK{{\cal K}}

\def\CN{{\cal N}}
\def\MeV{{\rm MeV}}
\def\bfq{{\bf q}}

\def\bfp{{\bf p}}

\def\too#1{\,\mathop{\longrightarrow}\limits_{#1}\, }
\def\bar#1{\overline{#1}}

\def\bra#1{\left\langle #1\right|}
\def\ket#1{\left| #1\right\rangle}

\def\half{{\textstyle{1\over2}}} %puts a small half in a displayed eqn
\def\frac#1#2{{\textstyle{#1\over #2}}}

\def\msb{{\mathop{{\bar {MS}}}}}
%
%       relations
%
\def\ltap{\ \raise.3ex\hbox{$<$\kern-.75em\lower1ex\hbox{$\sim$}}\ }
\def\gtap{\ \raise.3ex\hbox{$>$\kern-.75em\lower1ex\hbox{$\sim$}}\ }
\def\gl{\ \raise.5ex\hbox{$>$}\kern-.8em\lower.5ex\hbox{$<$}\ }
\def\roughly#1{\raise.3ex\hbox{$#1$\kern-.75em\lower1ex\hbox{$\sim$}}}

\def\eg{\hbox{\it e.g.}}

\def\np#1#2#3{{Nucl. Phys. } B{#1} (#2) #3}
\def\pl#1#2#3{{Phys. Lett. } {#1}B (#2) #3}
\def\prl#1#2#3{{Phys. Rev. Lett. } {#1} (#2) #3}
\def\physrev#1#2#3{{Phys. Rev. } {#1} (#2) #3}

\relax
\def\Dsl{\,\raise.15ex \hbox{/}\mkern-13.5mu D}
\def\[{\left[}
\def\]{\right]}
\def\({\left(}
\def\){\right)}
\noblackbox
%\draftmode
\def\pyidk{PHY-9057135}
\font\ninerm=cmr9
\def\Title#1#2{\nopagenumbers\abstractfont\hsize=\hstitle\rightline{{\ninerm
#1}}
\vskip .4in\centerline{\titlefont #2}\abstractfont\vskip .5in\pageno=0}
\Title{
\vbox{
\hbox{DOE/ER/40561-296-INT96-00-153 }
\hbox{UW/PT-96-31 }
\hbox{nucl-th/9610052}}}
{\vbox{\centerline{More Effective Field Theory}
\bigskip
\centerline{for Nonrelativistic Scattering }
\bigskip\bigskip
%\centerline{\bf PRELIMINARY DRAFT}
}}
\vskip-.2in
\centerline{David B. Kaplan}
\centerline{{\sl
Institute for Nuclear Theory, University of Washington}}
\centerline{{\sl Box 351550, Seattle WA 98195-1550}}
\centerline{{\tt dbkaplan@phys.washington.edu}}
\vfill
{An effective field theory treatment of nucleon-nucleon scattering at low energy shows much promise and could prove a useful tool in the study of nuclear matter at both ordinary and extreme densities.  The analysis is complicated by the existence a large length scale --- the scattering length --- which arises due to couplings in the short distance theory being near critical values.  I show how this can be dealt with by introducing an explicit $s$-channel state  in the effective field theory. The procedure is worked out analytically in a toy example.   I then demonstrate that a simple effective field theory  excellently reproduces the ${ }^1S_0$  $np$ phase shift up to the pion production threshold. }
\Date{10/96}
\baselineskip 18pt

\newsec{Introduction}

Effective field theory is a  powerful tool in particle physics for describing a scattering amplitude in a physical system with several length scales (for recent reviews, see \ref\eftrev{
H. Georgi, Ann. Rev. of Nucl. and Part. Sci., 43 (1993) 209;
J. Polchinski, Proceedings of  {\it Recent Directions in Particle Theory} ,
TASI92
(1992) 235;
A. V. Manohar, {\it Effective Field Theories}, hep-ph/9508245;
D.B. Kaplan,  {\it Effective Field Theories}, hep-ph/9506035.
}). 
It has been used with great success to describe both relativistic systems (\eg, the Standard Model of weak interactions) and nonrelativistic ones (\eg, positronium \ref\hqet{W.E. Craswell, G.P. Lepage, \pl{167}{1986}{437}}, as well as bound states containing heavy quarks \nref\nrqcd{G.T. Bodwin, E. Braaten, G.P. Lepage, \pl{1986}{167}{437}}\nref\hqet{H. Georgi, \pl{240}{1990}{447}; E. Eichten, B. Hill, \pl{243}{1990}{427}; N. Isgur, M.B. Wise, \pl{232}{1989}{113}, \pl{237}{1990}{527}}\refs{\nrqcd,\hqet}).  Such applications  involve a systematic expansion  in external momenta that allows one to compute low energy physical observables consistently to a desired accuracy.  Recently a number of papers have applied effective field theory to nucleon-nucleon interactions, with an eye toward better understanding the properties of nuclei and nuclear matter \hbox{\nref\weinberg{
 S. Weinberg, \pl{251}{1990}{288};
\np{363}{1991}{3}; \pl{295}{1992}{114}.}\nref\kolcka{
C. Ordonez, U. van Kolck, \pl{291}{1992}{459};
C. Ordonez, L. Ray, U. van Kolck, \prl{72}{1994}{1982};
Phys. Rev. C 53 (1996) 2086, nucl-th/9511380.}\nref\kolckb{ U. van Kolck, \physrev{C49}{1994}{2932}.}\nref\ksw{D.B. Kaplan, M.J. Savage, M.B. Wise, {\it Nucleon-Nucleon Scattering from Effective Field Theory}, nucl-th/9605002.  To appear in Nucl. Phys. B.}\refs{\weinberg-\ksw}}.
In this case one is dealing with nonrelativistic scattering and a short-range interaction, but the power expansion of the effective theory is complicated by the presence of an independent length scale in the form of  large scattering lengths.  For example, $np$ scattering in the ${ }^1S_0$ channel has a scattering length  $a=-23{\rm fm} = -1/(8 \MeV)$. In this letter I show how to construct an effective theory for such a system.  I demonstrate the method in an analytically soluble model, and then apply it to the problem of ${ }^1S_0$ $np$ scattering.  I find that the method, which includes in the effective theory a fundamental field exchanged in the $s$-channel, describes $np$ scattering far more successfully than previous effective field theory calculations (see fig.~5 below).

 To motivate what follows, I briefly review the justification for effective field theory. Experiments at momenta $p\ll \Lambda$ are insensitive to the effects of cuts and poles in the amplitude associated with the heavy states at $p\gtap \Lambda$.  Therefore  the amplitude can be well approximated by a function which correctly reproduces the nonanalytic features associated with light states at scales $p\ll \Lambda$, plus a smoothly varying function which can be expanded in powers of $p/\Lambda$ to account for the effects of the distant heavy states. This can be achieved by introducing an effective Lagrangian $\CL_{\rm eff}$  that includes the light states as propagating  degrees of freedom, along with nonrenormalizable local interactions in a power expansion in derivatives and fields suppressed by the appropriate powers of $\Lambda$.   The masses and couplings of the light states are adjusted to put the low lying singularities of the amplitude in the right places.  

If  the correct short distance theory is known, this matching of low lying singularities can sometimes be done by explicitly  by ``integrating out'' the heavy degrees of freedom.  In this case, the effective theory is a tool that greatly simplifies calculations by eliminating degrees of freedom that decouple from low energy phenomenology.  

On the other hand, if the correct short distance theory is not known, or the matching cannot be performed, then the couplings in the low energy theory are determined from data.  An example of this ``bottom-up'' effective theory is the chiral Lagrangians for pion physics.  Such theories typically are able to describe low energy physics in terms of few parameters, and are very predictive. 

This letter is organized as follows: I first perform a top-down calculation for a soluble system  with short distance interactions (the three dimensional $\delta$-shell potential),  performing the nonperturbative matching onto the effective theory analytically, showing how to avoid problems as the interaction is tuned such that the scattering length diverges.  I then generalize the matching procedure to a system with both short and long distance interactions. I then apply the method with great success to $np$ scattering in the ${}^1S_0$ channel.  The latter is a bottom-up calculation, where the coefficients of the effective theory are chosen to best fit the data. For simplicity I restrict my discussion to $s$-wave scattering throughout the paper.

\newsec{A toy model: the $\delta$-shell potential}

It is instructive to examine how the effective field theory works for an analytically soluble short distance interaction.  The example I use is of two ``nucleons'' of mass $M$ interacting in three dimensions via the potential $V(r)$ given by
\eqn\pot{V(r) = - g {\Lambda\over M} \delta(r-1/\Lambda)\ ,}
where $r=\vert {\bf r}_1 - {\bf r}_2\vert$ is the internucleon separation.  The parameter $\Lambda$ has dimensions of mass (I set $c=\hbar=1$) and sets the size of the interaction region. The coupling $g$ is dimensionless, and for $s$-wave scattering, there is a single boundstate for $g\ge 1$ and no boundstate for $g<1$.  

In the center of mass frame of the two nucleons, the $s$-wave Feynman scattering amplitude $\CA= -iT$ is given in terms of  
$\xi\equiv p/\Lambda = \sqrt{ME}/\Lambda$ and the spherical Bessel functions $j_0$ and $n_0$ \ref\gott{For a complete treatment of this potential, see: K. Gottfried, {\it Quantum Mechanics: Fundamentals}, Adison-Wesley Pub. Co. (1989)}:
\eqn\dsamp{\CA ={4\pi\over \Lambda M}\({g [j_0(\xi)]^2\over1 +g\xi n_0(\xi) j_0(\xi) -ig \xi [j_0(\xi)]^2}\)\ .}
This amplitude has a simple pole at
\eqn\pole{\xi_0 = i\eta, \ \ {\rm where}\ \ {1-{\rm e}^{-2\eta}\over 2\eta } = {1\over g}\ ;}
and $\vert \xi_0\vert \to 0$ as $g\to 1$.  This singular behavior is evident from the scattering length $a$  computed from \dsamp.  One finds
\eqn\scattlength{a = -\({g\over 1-g}\) {1\over \Lambda}\ ,}
which diverges as $g\to 1$.  Evidently a 
naive expansion of the amplitude  about $p\ll \Lambda$ (small $\xi$) will fail for $ \vert\xi\vert \gtap \vert (1-g)/g\vert$.  It is the dimensionless factor of $g/(1-g)$ that makes simple dimensional analysis for this system unreliable for $g\sim 1$.

Nevertheless, all other poles in the amplitude \dsamp\ occur at values   $\vert\xi\vert> 2\pi$. 
Following the precepts of effective field theory, the reasonable approach is to describe this system as one of nucleons $N$,  a fundamental field $\phi$ which has baryon number $B=2$ and mass $2M+\Delta$ (henceforth called ``the dibaryon''), and local interactions expanded in powers of $\nabla/\Lambda$. The field $\phi$ is included to reproduce the low-lying pole in the amplitude; a radius of convergence of $\vert p\vert \gtap \Lambda$ is expected for the residual interactions, since all other singularities in the amplitude are at that scale. The effective Lagrangian is given by
\eqn\leff{\eqalign{
\CL_{\rm eff} = &N^{\dagger} \(i\partial_t-\nabla^2/2M\) N +\sigma\phi^\dagger\(i\partial_t-\nabla^2/4M - \Delta\)\phi\cr  &-(y\phi^*NN +{\rm h.c.}) -\half C(N^\dagger N)^2+\ldots}}
where the ellipses refer to higher dimension operators suppressed by powers of $\Lambda$.  The field $\phi$ can be rescaled so that $\vert \sigma\vert=1$, but the sign of $\sigma$ will have to be determined. In the center of mass frame, the scattering amplitude can be calculated by summing the ladder diagrams (solving the Lippman-Schwinger equation) with the kernel 
\eqn\veff{ V_{\rm eff}(p) = C + {y^2\over \sigma E-\Delta} \ ,\qquad E=p^2/M,}
as shown graphically in fig.~1.  
The bubble diagrams in the geometric sum are divergent, but renormalization of $C$, $\Delta$ and $y$ renders the theory finite.  Of course, the values for these renormalized parameters are scheme dependent and not physically meaningful.  What is desired is a scheme that facilitates the power counting and effective field theory expansion in $p/\Lambda$.  Dimensional regularization and minimal subtraction ($\bar{MS}$) is such a scheme, as the renormalization scale $\mu$ only appears in logarithms;  I use this scheme throughout the paper.  The basic calculation one needs for the present model is the bubble integral \ksw
\eqn\dimreg{
 \int {{\rm d}^{3} \bfq\over (2\pi)^{3}}\,{1\over (E -
{\bf q}^2/M+i\epsilon)}\longrightarrow {M\sqrt{-ME -i\epsilon}\over 4 \pi}=
{-iM|\bfp |\over 4\pi}}
The resultant amplitude is
\eqn\ampeff{\CA_{\rm eff} = -{V(p)\over 1 + i M V(p)p/4\pi} = -{C(\sigma E-\Delta)+y^2\over \sigma E-\Delta +i(M p/4\pi)[C(\sigma E-\Delta)+y^2]}\ .} 
\topinsert
\centerline{\epsfxsize=3in\epsfbox{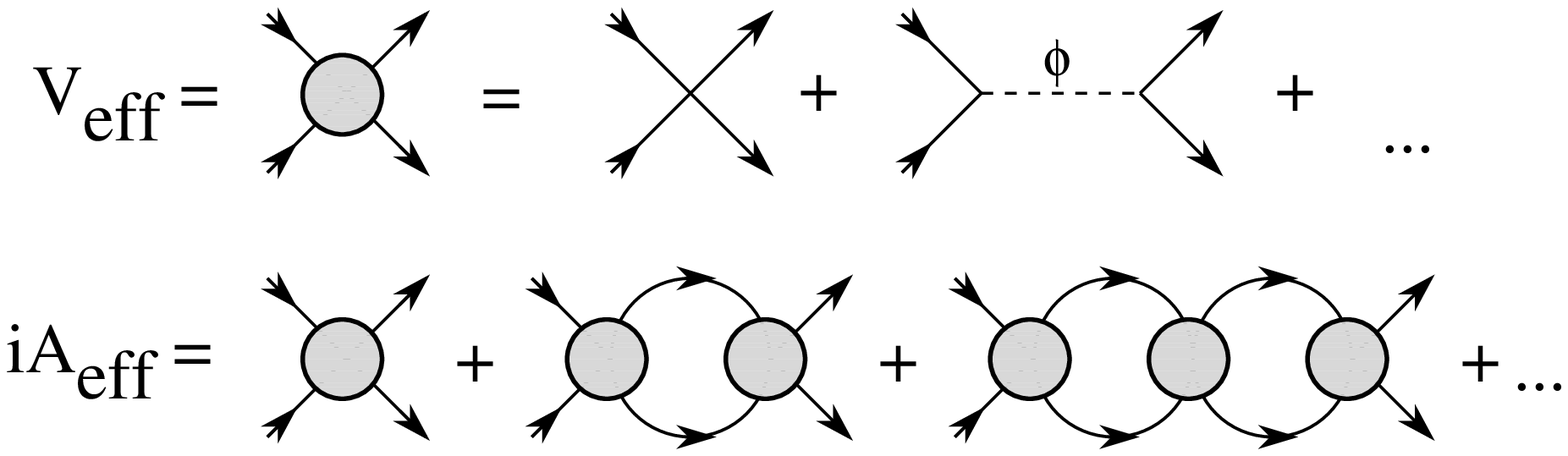}}
\smallskip
\caption{Fig.~1.  The potential  from the effective theory \leff, and the amplitude \ampeff\ obtained from summing the ladder (bubble) diagrams.}
\endinsert

In eqs. \dsamp\ and \ampeff\ I have written the exact and effective amplitudes in the form $N(p)/D(p)$ where $N$ and $D$ are nonsingular functions of $p$.  By correctly choosing the parameters of the effective theory, it is possible to match $N$ and $D$ in the two theories up to $\CO(p^4)$ corrections  and  a common multiplicative constant which factors out of the amplitude.  
This involves solving the following equations to order $\CO(p^2)$:
\eqn\matchit{\eqalign{
-&[C(\sigma E-\Delta)+y^2]= z {4\pi\over \Lambda M} g [j_0(\xi)]^2\ ,\cr
 &(\sigma E-\Delta) = z\(1+ g \xi j_0(\xi)n_0(\xi)\)\ ,}}
where $z$ is the irrelevant constant.

One finds the following solution to the matching equations \matchit:
\eqn\params{\eqalign{ 
\Delta &= -\sigma \({1-g\over g}\){3 \Lambda^2\over 2 M} = \sigma {3\Lambda\over a M}\ ,\cr
C&= {2\pi\over M\Lambda}\ ,}
\qquad\qquad
{\openup 1\jot\eqalign{y^2&=- \sigma \({1+g\over g}\) {3\pi \Lambda\over M^2}\ ,\cr
\sigma &=\pm 1\ .}}}
The sign of $\sigma$ is determined by the fact that $y^2$ must be positive:
\eqn\sigval{\sigma = -{\rm Sign}\[{1+g\over g}\]\ .}
There are several interesting features about these values:
\item{(i.)} The parameter $\Delta$ is proportional to the inverse scattering length $1/a$, and goes to zero at the critical coupling $g=1$.  Since the momentum variation of the $\phi$ propagator depends on the combination $p^2/M\Delta$, for $g\sim 1$ the $\phi$ field is  needed in the effective theory to reproduce the rapid variation of the scattering amplitude at low $p$.  On the other hand, for $\vert g/(1-g)\vert\sim \CO(1)$, then the scattering length is short and  the $\phi$ field looks heavy ($M\Delta\sim \Lambda^2$); it can be integrated out without compromising the convergence of the momentum expansion, but then the effective theory is simply perturbative \foot{For an enlightening discussion, see ref. \ref\luma{M. Luke, A.V.Manohar,  hep-ph/9610534}.}.  
\item{(ii.)} The parameter $C$ is independent of $g$ and is small, in the sense that $CMp/(4\pi) \sim p/\Lambda$, so that its effects only become nonperturbative for large momenta, $p\sim \Lambda$.
\item{(iii.)} For $\sigma=-1$, which is the case when \hbox{$\vert g+\half\vert > \half$}, the $\phi$ field in the effective theory \leff\  has the wrong-sign $\partial_t$ energy term.  This is no cause for alarm, since the nonperturbative amplitude one computes using the fully dressed $\phi$ propagator and vertices is unitary and sensible.  Loops with internal $\phi$ particles are not included in the effective theory. 
\topinsert
\centerline{\epsfxsize=2.5in\epsfbox{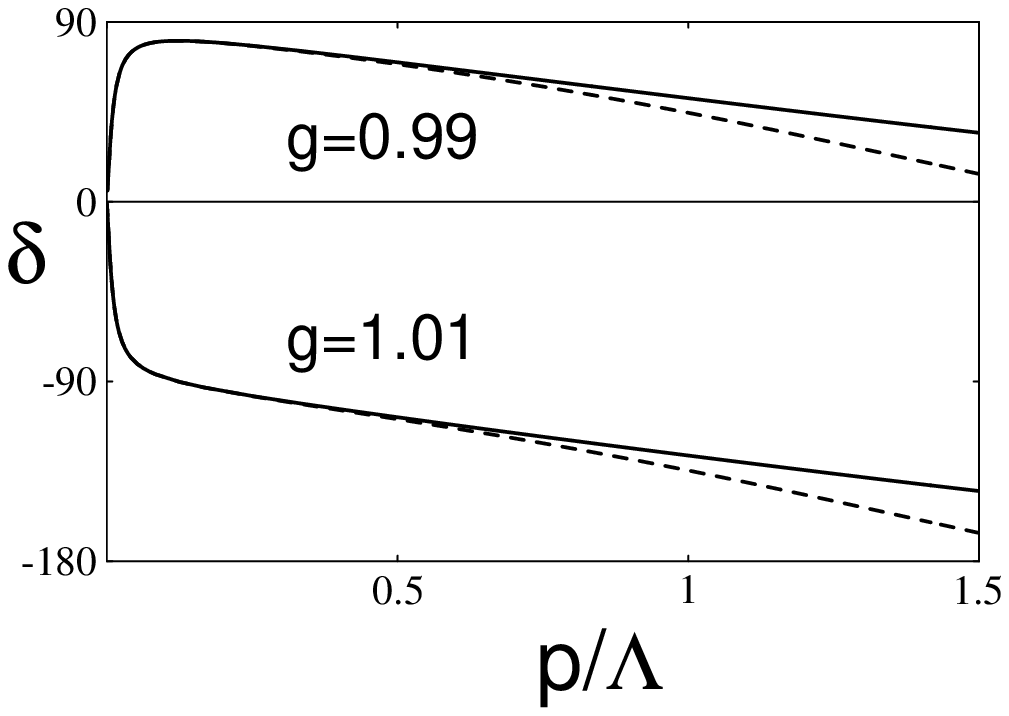}}
\smallskip
\caption{Fig.~2. $s$-wave phase shift (degrees) for the $\delta$-shell potential, plotted as a function of $p/\Lambda$ for $g=0.99$ (with an almost-bound state) and $g=1.01$ (with a weakly bound state).  The solid lines are the exact results, while the dashed lines are computed from the effective field theory as described in the text, with $M/\Lambda=1.2$.}
\endinsert

In fig.~2 I compare the exact phase shift from eq. \dsamp\ with the effective field theory prediction \ampeff\ for the two values $g=1\pm .01$.  These are systems where the scattering length $a$ in eq. \scattlength\ is 100 times the fundamental length scale  of the problem, $1/\Lambda$.   As one can see, the size of the factor $g/(1-g)$ does not affect the range of validity of the effective field theory.  
In fig.~3 I do the same for very strong couplings $g=\pm 10$.  As expected, in every case one compares, whether it be strong or weak coupling, attractive or repulsive, the effective theory agrees well with the exact result for momenta $p\ltap \Lambda$.  Evidently the effective theory \leff\ is very effective at  accurately describing long distance physics.

\topinsert
\centerline{\epsfxsize=2.5in\epsfbox{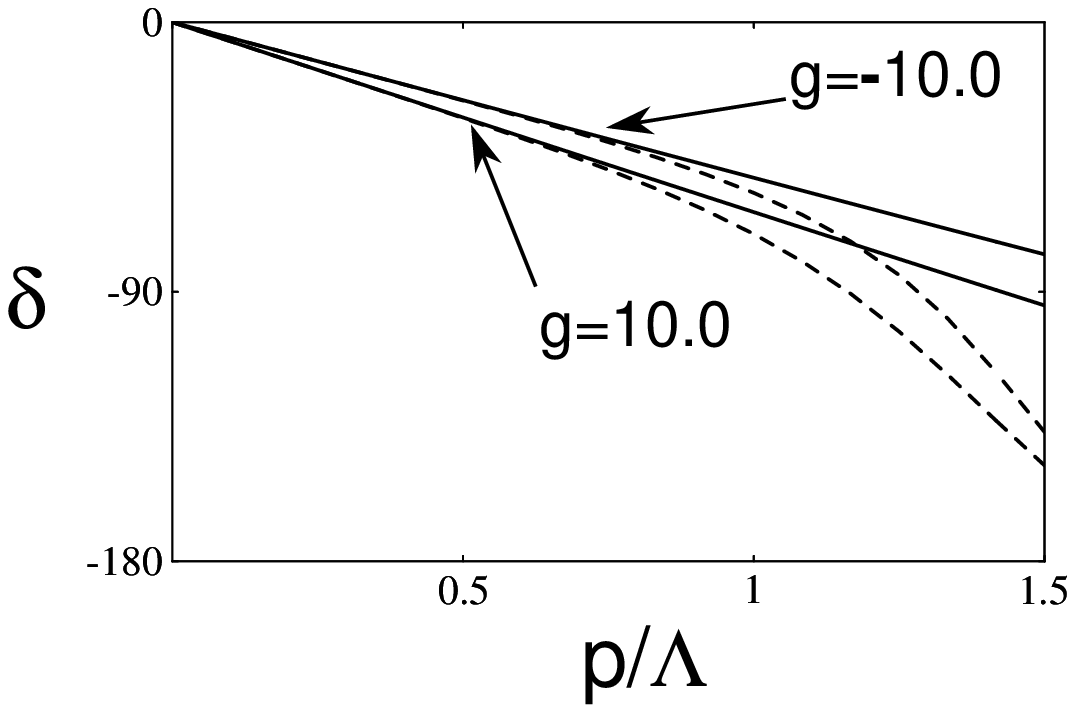}}
\smallskip
\caption{Fig.~3. $s$-wave phase shift (degrees) for the $\delta$-shell potential, for $g=10.0$ (strongly attractive) and $g=-10.0$ (strongly repulsive).  The solid lines are the exact results, while the dashed lines are computed from the effective field theory with $M/\Lambda=1.2$.}
\endinsert

If  one tries to do without the dibaryon state $\phi$ in the effective theory, using instead the effective Lagrangian
\eqn\nophi{\tilde\CL_{\rm eff} = N^{\dagger} \(i\partial_t-\nabla^2/2M\) N  -\half \tilde C(N^\dagger N)^2- \half\tilde C_2 (N^\dagger\nabla N)^2+\ldots}
then one runs into difficulties.  The couplings of this theory $\tilde\CL_{\rm eff}$  can be readily obtained from $\CL_{eff}$ in eq. \leff\ by integrating out the $\phi$ field.  This amounts to performing a Taylor expansion of interaction $V_{\rm eff}(p)$ in eq. \veff\ for small $p$.  One finds  $\tilde C=4\pi a/M$ (as was found in ref. \weinberg), and that higher derivative operators in $\tilde \CL_{eff}$ form an expansion in powers of $ p^2/M\Delta \sim p^2 a/\Lambda$.   This expansion is fine so long as there is no boundstate near threshold.  However, when $a$ is large, $\Delta$ is small and  $\tilde \CL_{\rm eff}$ does not constitute a useful expansion. Such is the case for realistic $s$-wave $NN$ scattering, for which the scattering lengths are much larger than the range of the short distance interaction.

This observation exemplifies the general discussion of ref. \ksw: that the radius of convergence of the derivative expansion an effective field theory without a dibaryon is set by the scale $p\sim \sqrt{\Lambda/a}$, which is poor for systems with large scattering length $a$.  However it was argued in \ksw\ that in fact the higher derivative operators were all correlated, and that the expansion of $1/\CA$ would have a radius of convergence  $p\sim 1/\Lambda$.  The  method 
described here of including an explicit dibaryon field in the $s$-channel is a more straightforward and intelligible procedure containing the same physics.

%%%%%%
\newsec{A toy model with both short- and long-range interactions}

Performing an effective field theory calculation for nucleon interactions is complicated by the existence of both short- and long-range  interactions.  The latter is due to one pion exchange (OPE), which must be included explicitly in the effective theory. One might worry that the effects of pion interactions could completely change the nature of the solution \params, particularly if pion exchange is strong enough to contribute significantly to the total scattering length. It is therefore instructive to complicate the $\delta$-shell model of the previous section by including a Yukawa interaction with range $1/m_\pi\gg 1/\Lambda$.  Using the techniques developed in ref. \ksw, it is possible to perform the matching analytically.  In fact, I will show that even if the pion interaction is strong, its effects on the matching coefficients is computable and small --- $\CO({m_\pi\over \Lambda})$. The reader uninterested in technical details can skip to \S4.

\subsec{The  amplitude of the full theory}
For the interaction in the full theory I take the potential $V(r)$, where 
\eqn\fullv{
V = V_\delta + V_\pi\ ,\qquad V_\delta= -{g\Lambda\over M}\delta(r-1/\Lambda)\ ,\qquad   V_\pi = - {\alpha_\pi\over r}{\rm e}^{-m_\pi r}\ .}
Following ref. \ksw, I define the OPE Hamiltonian $H_\pi = -\nabla^2/M + V_\pi$ with dimensionless solutions
$\CJ_p$, $\CN^\lambda_p$ satisfying \foot{Comparing with analogous functions defined in ref. \ksw, $\CJ_p(r)=\CJ_E(r)$ and  $\CN_p^\lambda(r)= -(4\pi/Mp)\CK^\lambda_E(r)$, where $p=\sqrt{EM}$.}
\eqn\hpisol{
\(H_\pi - E\) \CJ_p(r) = 0\ ,\qquad  \(H_\pi - E\) \CN^\lambda_p(r) = -{4\pi\over M p}\delta^3({\bf r})\ ,}
with expansions near the origin 
\eqn\expan{\openup2\jot\eqalign{
\CJ_p(r) {\,\smash{\mathop{\longrightarrow}\limits_{r\to 0}}\,}& 1 -{\alpha_\pi M\over 2} r + {2\alpha_\pi m_\pi M + \alpha_\pi^2 M^2 -2p^2\over 12} r^2 +\CO(r^3)\ ,\cr
\CN^\lambda_p(r){\,\smash{\mathop{\longrightarrow}\limits_{r\to 0}}\,}  &-{1\over p}\biggl( {1/ r} - \alpha_\pi M\ln\lambda r + {\alpha_\pi^2 M^2\over 2}r\ln\lambda r -  {3\alpha_\pi^2M^2 -2\alpha_\pi M m_\pi + 2p^2)\over 4} r \biggr. \cr  &\biggl. +\CO(r^2\ln \lambda r)\biggr)\ .}}
The parameter $\lambda$ is due to the ambiguity of being able to add the solution $\CJ$ to $\CN$ while still satisfying eq. \hpisol; nothing physical will depend on it. 

An $s$-wave  scattering solution with momentum ${ p}$ to the Schr\"odinger equation with the full potential $V$ given in eq. \fullv\  can be written as $\psi_{ p}(r)= a\CJ_p(r) + b\CN^\lambda_p(r)$ for $r>1/\Lambda$, while $\psi_{ p}(r)\propto \CJ_p(r)$ for $r<1/\Lambda$.  It is straightforward to compute the ratio $a/b$ by performing the matching at $r=1/\Lambda$, making use of the Wronskian\foot{$W = \CN_p^\lambda {{\rm d\ }\over {\rm d}r} \CJ_p  -
\CJ_p {{\rm d\ }\over {\rm d}r} \CN_p^\lambda 
 = -{1 \over p r^2}$}:
\eqn\abrat{{a\over b} = {M\over 4 \pi} \[{\CN^\lambda_p\over \CJ_p}p + {\Lambda\over  g (\CJ_p)^2}\]_{r=1/\Lambda}\ .}

The $s$-wave phase shift is then determined by $a/b$ and the asymptotic properties of $\CJ$ and $\CN$.  The latter can be expressed as
\eqn\asymprop{
\CJ_p(r)\too{r\to \infty}  \(\eta {e^{ipr}/ pr} + c.c.\)\ ,\qquad
\CN^\lambda_p(r) \too{r\to \infty}  \(\zeta {e^{ipr}/ pr} + c.c.\)\ .}
The phase shift is then given by
\eqn\phase{ {\rm e}^{2i\delta} = -{(a\eta+b\zeta)^{\ }\over (a\eta+b\zeta)^*}\ ,}
where $\eta,\zeta$ are the asymptotic coefficients of eq. \asymprop. 
Following ref. \ksw, it is convenient to use the expressions
\eqn\piexpress{\eqalign{
{\eta/ \eta^*}&= {\rm e}^{2i\delta_\pi}\ ,\cr
{1/ \eta^*} & = 2i\chi_{\bf p}({\bf 0})\ ,\cr
{\zeta^*/ \eta^*}&= G_E^{\msb} ({\bf 0,0})+    {\alpha_\pi
M^2\over 8\pi} \left[\ln {\mu^2\over\lambda^2} + 2\gamma -
1\right]\ .}}
In the above expression,  $\delta_\pi$,  $\chi_{\bf p}(0)$, $G_E^{\msb} ({\bf 0,0})$ are quantities computed from the OPE Hamiltonian $H_\pi$:
$\delta_\pi$ is the OPE phase shift, $\chi_{\bf p}(0)$ is the OPE wavefunction at the origin, and $G_E^{\msb} ({\bf 0,0})$, the Green function $\bra{\bf r} 1/(E-H_\pi)\ket{\bf r'}$ at ${\bf r}' = {\bf r}=0$, dimensionally regulated and renormalized in the $\bar{MS}$ scheme, at the renormalization scale $\mu$ \foot{The calculation of the amplitude $\CA$ never requires renormalization; I express the answer in terms of the renormalized propagator  to more easily compare to the effective theory, which does require renormalization.}.  

Combining eqs. \abrat-\piexpress\ I determine the complete scattering amplitude to be
\eqn\dpiamp{
\CA = \CA_\pi +   
\({4\pi \over M\Lambda}\){g \CJ_p^2  \(\chi_{\bf p}({\bf 0})\)^2 \over
1+g\xi \CN^\lambda_p \CJ_p + g {4\pi\over M \Lambda}\CJ_p^2  \[{\alpha_\pi
M^2\over 8\pi} \left(\ln {\mu^2\over\lambda^2} + 2\gamma -
1\right)
 +G_E^{\msb} ({\bf 0,0})\]}
.}
In this expression, $\CJ_p$  and   $\CN^\lambda_p$ are evaluated at $r=1/\Lambda$.

\def\atoz{\smash{\,\mathop{\longrightarrow}\limits_{\alpha_\pi\to 0}}\,}
Note that in the limit   that the pion coupling vanishes ($\alpha_\pi\to 0$) the terms in the above expression behave as 
\eqn\pilim{\CA_\pi\atoz 0\ ,\qquad \chi_{\bf p}({\bf 0})\atoz 1\ ,\qquad  
 G_E^{\msb} ({\bf 0,0})\atoz -{i M p\over 4\pi}\ ,}
while the $\CJ$ and $\CN$ functions become spherical Bessel functions:
\eqn\sbes{\CJ_p(r)\atoz j_0(p r)\ ,\qquad
\CN^\lambda_p(r) \atoz n_0(p r) \ .}
Upon substituting the above  expressions in eq. \dpiamp\ one recovers the amplitude \dsamp\ for the $\delta$-shell potential alone,  $\alpha_\pi=0$.

%%%%%%%%%
\subsec{The amplitude in the effective theory}

The effective theory with which I expect to describe low energy scattering consists of a contact interaction  and both $\phi$ and  $\pi$ exchange. The potential  for the effective theory is given graphically in fig.~4;
it is the same as the effective interaction in fig.~1, with the addition of the single pion exchange graph:
\eqn\nppot{V_{\rm eff} \equiv V_0 + V_\pi\ ,\qquad
V_0(E) = C  + {y^2 \over \sigma E - \Delta}\ ,\qquad
V_\pi({\bf p},{\bf p'})=  -{\alpha_\pi\over ({\bf p}-{\bf p'})^2+m_\pi^2}\ .}
Using the techniques of ref. \ksw\ it is possible to write the resultant scattering amplitude in closed form as 
\eqn\epiamp{\openup 2\jot\eqalign{\CA_{\rm eff} &= \CA_\pi -  {V_0(E)\(\chi_{\bf p}({\bf 0})\)^2 \over
1 - V_0(E) G_E^{\msb} ({\bf 0,0})}\cr
&= \CA_\pi 
-{\[C(\sigma E-\Delta)+y^2\] \(\chi_{\bf p}({\bf 0})\)^2 \over \sigma E-\Delta -G_E^{\msb} ({\bf 0,0})[C(\sigma E-\Delta)+y^2]}
\ .}}
%%%%%%%

%
\topinsert
\centerline{\epsfxsize=3in\epsfbox{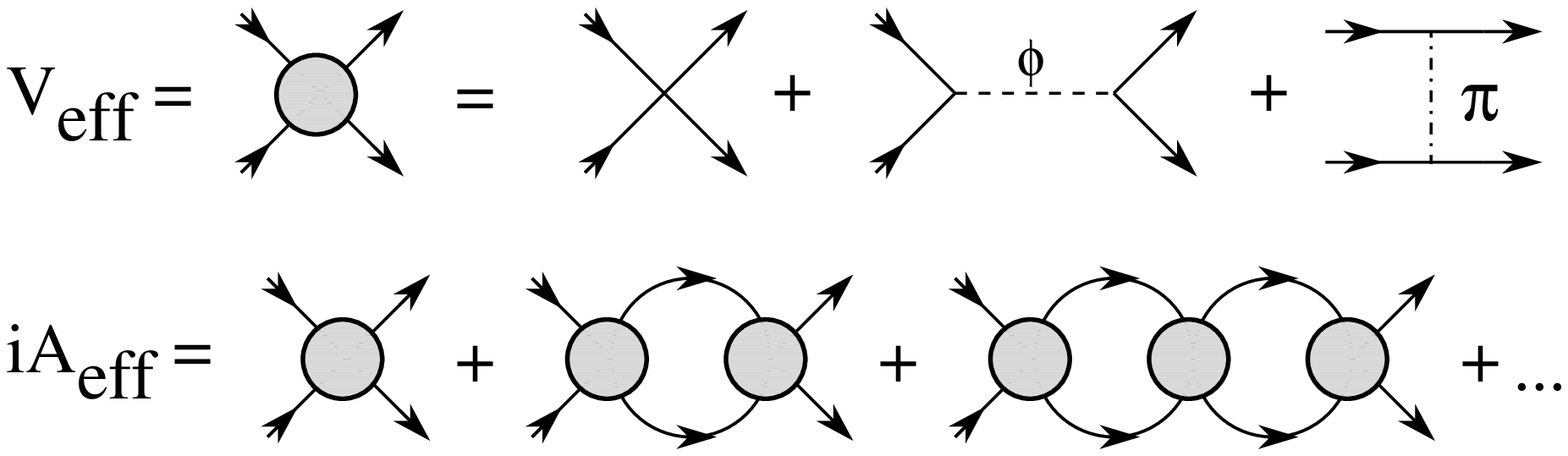}}
\smallskip
\caption{Fig.~4.  The potential  from the effective theory  including the explicit dibaryon field $\phi$ and single pion exchange.  The amplitude $\CA_{\rm eff}$ is obtained from summing the ladder (bubble) diagrams. Techniques for performing the sum using dimensional regularization are extensively discussed in ref. \ksw.}
\endinsert
\subsec{Matching the effective and full amplitudes}
Matching entails choosing the free parameters  $\{y^2,\ \Delta,\ C\}$ such that the amplitude \epiamp\ best approximates the exact exact amplitude  \dpiamp.  
The analogue of the matching equations \matchit\  in terms of $\xi=p/\Lambda$ are:
\eqn\matchitii{\openup 2\jot\eqalign{
-&[C(\sigma E-\Delta)+y^2]=z {4\pi\over \Lambda M} g [\CJ_p]^2\ ,\cr
 &(\sigma E-\Delta) = z\[1+ g \xi \CJ_p\CN^\lambda_p+
g {\alpha_\pi M\over 2\Lambda} \CJ_p^2   \left(\ln {\mu^2\over\lambda^2} + 2\gamma -
1\right)  
\]\ ,}}
where again $\CJ_p$ and $\CN^\lambda_p$ are evaluated at $r=1/\Lambda$. These matching conditions are to be solved to $\CO(p^2)$. 

So what has been the effect of the pions?  The effective theory is supposed to be an expansion in $p/\Lambda$; must one perform a simultaneous expansion in $\alpha_\pi$?  To resolve this question note that one pion exchange must be treated nonperturbatively for $\alpha_\pi M/m_\pi\gtap 1$.   In the real world, this value is nearly realized --- for $NN$ scattering in the ${ }^1S_0$ channel, $\alpha_\pi = 0.47 m_\pi/M$. This suggests that one should consider $\alpha_\pi \sim m_\pi/M$ in any power counting scheme.  Furthermore, for momenta $ m_\pi\ltap p\ll \Lambda$,   the effective theory is an expansion in $\epsilon = p/\Lambda \gtap m_\pi/\Lambda$. Combined, these considerations imply
\eqn\powerc{\openup 1\jot\eqalign{ &{p\over \Lambda} \sim {m_\pi\over \Lambda} \sim {\alpha_\pi M\over \Lambda}\sim \epsilon \cr
&{\alpha_\pi M m_\pi \over \Lambda^2} \sim {\alpha_\pi^2 M^2\over \Lambda^2}\sim \epsilon^2\ .}}
Thus when $\CJ_p(r)$ is evaluated at $r=1/\Lambda$ for the matching condition \matchitii, I find from eq. \expan\ 
\eqn\expanii{\CJ_p(1/\Lambda) = \(1+\CO(\epsilon)\)  -{p^2\over 6\Lambda^2}\(1+\CO(\epsilon)\)\ .}

An important lesson has been learned from this exercise:   one can  perform the matching in a perturbative expansion in $\epsilon\sim m_\pi/\Lambda$, even when pion interactions are nonperturbative and $\alpha_\pi M/m_\pi\sim 1$. Since $\alpha_\pi$ enters the matching via the combinations in eq. \powerc, the leading order calculation in $\epsilon$ corresponds to setting $\alpha_\pi$ to zero, and one recovers the solution \params\ found when performing the matching without pions. At subleading order, the $\CO(\epsilon)$ contributions from matching are more important than including two derivative operators in $\CL_{\rm eff}$, which are $\CO(\epsilon^2)$ effects. One may be tempted to think that this is a manifestation of chiral symmetry, since $\alpha_\pi \propto m_\pi^2/f_\pi^2$ and vanishes in the chiral limit.  However, this is {\it not} the case, since we are assuming that $\alpha_\pi M/m_\pi \sim \CO(1)$, while that combination would vanish in the chiral limit.

Finally note that the $\ln {\lambda\over\Lambda}$ terms in $\CN^\lambda_p$ combine with the $\ln {\mu^2\over \lambda^2}$ term in eq. \matchitii\ to give $\ln {\mu^2\over \Lambda^2}$.  This logarithm accounts for running of the effective interactions at scales $\mu$ below the matching scale $\Lambda$.

%%%%%%%%%%%%%
\newsec{$NN$ scattering in the ${ }^1S_0$ channel}
\nref\nij{Partial wave analysis of the Nijmegen University theoretical high
energy physics group, obtained from WorldWide Web page
http://nn-online.sci.kun.nl/ .   For analysis see V.G.J. Stoks, R.A.M. Klomp, M.C.M. Rentmeester, and J.J. de Swart.   Phys.Rev.C48 (1993) 792}
I now turn to a problem of phenomenological interest: $np$ scattering.  As discussed above, this is a system with a scattering length $a=-23\ {\rm fm}$ much larger than the expected scale of short-range strong interactions (which are characterized by a scale shorter than 1 fm).  Therefore an effective field theory description can be expected to benefit by explicit inclusion of a dibaryon field $\phi$ as discussed in the previous sections. The effective theory for $np$ must also include explicit pion fields.  Since the short-distance theory is not understood, I fit the couplings of the effective theory to data, rather than computing them via matching conditions.  The results are expected to be similar to those in ref. \ksw, where the $\phi$ field was not included explicitly, but where $1/\CA$ was expanded so as to be insensitive to the $\phi$ pole.  One advantage of the present method is that  unlike the treatment in \ksw, the present method should be valid even for small phase shift (where $1/\CA$ blows up). Furthermore, it is possible to account for the residual interactions in an expansion in $p/\Lambda$, where $\Lambda$ characterized short distance physics.

The potential  summed for $np$ scattering is given graphically in fig.~4  with \eqn\nppot{\eqalign{V_{\rm eff} &\equiv V_0 + V_\pi\ ,\cr
V_0(E)&= \(C+ {g_A^2\over 2 f_\pi^2}\) + {y^2 \over \sigma E - \Delta}\ ,\cr
V_\pi({\bf p},{\bf p'})&=  -{4\pi\alpha_\pi\over ({\bf p}-{\bf p'})^2+m_\pi^2}\ ,\qquad \alpha_\pi ={g_A^2 m_\pi^2\over 8\pi f_\pi^2}\ .}}
The pion axial coupling $g_A=1.25$ and decay constant $f_\pi=132\ \MeV$ are known; thus the effective theory involves the same three unknown parameters as in the $\delta$-shell example:  The four-nucleon interaction $C$, the dibaryon mass shift $\Delta$, and the dibaryon coupling $y$ (as well as the sign $\sigma$ of the $\partial_t$ kinetic term).  As before, the values for these quantities are not physical, depending on the renormalization scheme chosen.  The scattering amplitude one derives, however, is independent of the renormalization scheme.

The amplitude $\CA_{\rm eff}$ of the effective theory is the same as appearing in eq. \epiamp, with $V_0$ and $\alpha_\pi$ given by eq. \nppot. By numerically computing the quantities $\delta_\pi$, $\chi$ and $G$, and subsequently fitting the resultant phase shift to the data, I find  the following (scheme dependent) values for the unknown parameters at renormalization scale $\mu=m_\pi$:
\eqn\npparams{\eqalign{ 
C&= -\({1\over 192\ \MeV}\)^2\ , \cr
y^2 &= {1\over 507\ \MeV}\ ,
}
\qquad\qquad{\openup2\jot\eqalign{
\bar C &\equiv \(C+{g_A^2\over 2 f_\pi^2}\) = \({1\over 237\ \MeV}\)^2\ ,\cr
\Delta &= 14.8\ \MeV
\ ,}}}
and $\sigma=-1$.  Note that the contact interaction can be written as $\bar C = 4\pi/M\Lambda$, where $\Lambda = 749\ \MeV$; thus a loop with a contact interaction brings with it a power of ${\bar C M p\over 4\pi} = p/\Lambda$.  
As this is independent of $M$, the power counting scheme is different in such a theory from that described in refs. \refs{\weinberg-\ksw}; all residual interactions in the derivative expansion of $\CL_{\rm eff}$ can be treated perturbatively.
\topinsert
\centerline{\epsfxsize=2.5in\epsfbox{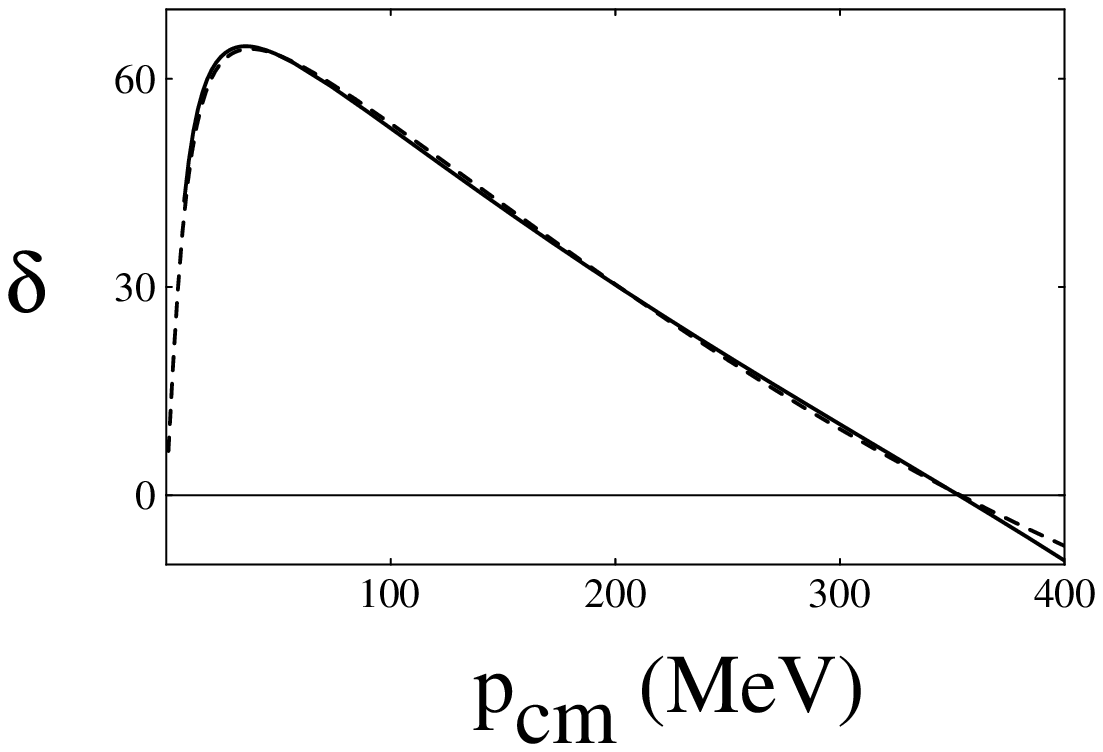}}
\smallskip
\caption{Fig.~5. ${ }^1S_0$ phase shift (degrees) versus center of mass momentum $p_{\rm cm}$ for $n-p$ scattering. ($p_{cm}=400\ \MeV$ corresponds to $T_{\rm lab}= 340\ \MeV$).  The solid line is taken from the Nijmegen partial wave analysis \nij, while the dashed line is computed  from the effective field theory in eq. \epiamp, which involves three free parameters.}
\endinsert
As is shown in fig.~5, the phase shift $\delta_{\rm eff}$ computed from this effective theory with its three free parameters is essentially indistinguishable from the results of the Nijmegen phase shift analysis for ${ }^1S_0$ scattering, all the way up to center of mass momentum $p_{\rm cm}=400\ \MeV$, corresponding to $T_{\rm lab} = 340\ \MeV$.  That the fit works so well is evidence that higher derivative are suppressed by powers of a large scale, while the relevant effects of multipion exchange, $\Delta$ exchange, {\it etc.} are well accounted for by the effective interactions  included in $\CL_{\rm eff}$ in eq. \nppot.

\newsec{Conclusions}
Expanding on techniques introduced in ref. \ksw, I have demonstrated that effective field theory and dimensional regularization can be successfully applied to the study of nonrelativistic systems with short-range interactions.  I have shown that when the short-range interaction is near the critical value for which the scattering length diverges, then a low-lying $s$-channel state must be introduced as a physical degree of freedom in the effective theory.   This procedure is in keeping with how effective field theory is applied to relativistic systems when a mass independent renormalization scheme is employed \eftrev.
As shown by explicit example, the method works well for  both attractive and repulsive interactions, despite the fact that problems can arise when using other forms of regularization \ref\cohen{T. D. Cohen,  nucl-th/9606044; D. R. Phillips, T.D. Cohen, nucl-th/9607048}.  

The effective theory reproduces the nonperturbative features of  interactions at the scale $\Lambda$  through the  $s$-channel resonance, plus local interactions which can be treated perturbatively in an expansion in $p/\Lambda$. Including additional long-range, nonperturbative interactions   does not disturb this power counting.

The method also works very well in reproducing $np$ scattering data in the ${ }^1S_0$ channel.  It is hoped that the technique can be used for the spin triplet channels as well, but that has yet to be resolved. Eventually, the value of the approach outlined here is hoped to lie in the study of nuclear matter at both normal and extreme density. While the usefulness of effective field theory has yet to be determined for bulk matter, it has been shown here to reproduce the two particle scattering amplitude at well above the nucleon Fermi momentum in matter, so there is reason for optimism.

\vfill
\centerline{Acknowledgements}

I wish to thank L. Brown, H. Georgi, P. Lepage, M. Savage, U. van Kolck and  M. Wise for useful conversations. This work was supported in part by DOE grant
DOE-ER-40561, and NSF Presidential Young Investigator award
\pyidk.

\listrefs
\bye